# Optimizing $\tilde{X}_W^{\pm} \rightarrow h^0 \ell_i^{\pm}$ Reconstruction Efficiency with Small-R (R = 0.4) and Large-R (R = 1.0) Jets

Sophie Kadan[1], Evelyn Thomson[2]

University of Pennsylvania[1,2], sokadan@sas.upenn.edu[1]

**ABSTRACT**

Many beyond the Standard Model searches by the ATLAS experiment at the CERN Large Hadron Collider employ jets to simplify event reconstruction. Two types of jets have been studied to optimize the selection and reconstruction of a final state with many jets originating from four b-quarks. A large-R (R = 1.0) jet combines particle shower products into one jet that spans 2 radians, while a small-R (R = 0.4) jet gives finer-grained information. In this talk, we consider the pair production of charginos with R-parity violating (RPV) decays to a charged lepton and a Higgs boson, which subsequently decays to two b quarks. In studies of Monte Carlo simulation, we found that parameters such as the distance between Higgs bosons and the distance between b-jets were relevant in selecting the most accurate small-R jet reconstructions. These parameters were then used to refine small-R jet selection, increasing reconstruction efficiency across a wide range of possible chargino masses compared to a selection using large-R jets. To extend these findings to a model with a RPV decay directly to two b quarks and a lepton without the intermediate Higgs boson, machine learning techniques are now being explored.

**CONFLICTS OF INTEREST**

On behalf of all authors, the corresponding author states that there is no conflict of interest.

# 1 Introduction

## 1.1 The Standard Model (SM)

The Standard Model (SM) of particle physics describes all known particles and their interactions. This model predicts matter to be fundamentally composed of generations of quarks and leptons, characterized by mass, electric charge, and spin. SM particles interact via the exchange of gauge bosons, which act as force carriers [1, 2]. Specifically, the strong, weak, and electromagnetic forces arise from these elementary gauge bosons—encompassing much of physics as we know it [2].

The recent discovery of the Higgs boson has further unified the electromagnetic and weak forces, introducing a source of elementary particle mass. Specifically, this mass is derived via Higgs field interactions, whose strength determines fundamental properties such as atom size and proton stability. Furthermore, interactions with the scalar Higgs field are consistent with experimental observations: they predict that W and Z bosons carry mass while photons and gluons remain massless. Uniting particles and their forces, the Higgs boson discovery thus completes the Standard Model [3].

While its predicted particles have all been experimentally confirmed, the Standard Model nonetheless has many shortcomings. For example, SM predictions do not provide a quantum description of gravity, the fourth fundamental force [4, 5]. Likewise, the Standard Model does not account for the low mass of the Higgs boson: termed the Hierarchy Problem, SM predictions overestimate the Higgs mass due to its interactions with virtual particles. Corrections must thus be made at scales where new physics is relevant [6].

## 1.2 Beyond the Standard Model (BSM)

To reconcile SM issues, many Beyond the Standard Model (BSM) theories have been developed. One such theory is Supersymmetry (SUSY), which introduces a "superpartner" for each currently known particle [4]. This theory maps particles of spin one half to those of integer spin and vice versa, establishing a symmetry that connects fermions and bosons [7]. Because superpartners of equal mass to their partners have not yet been discovered, supersymmetry is presumed to be a broken symmetry. To reduce potentially incorrect assumptions, the Minimal Supersymmetric Standard Model (MSSM) has been established as a SUSY model that only considers the minimal number of new particles [8].

SUSY is attractive in its potential ability to solve the Hierarchy Problem via loop corrections that its superpartners provide [5, 8]. Local SUSY also necessarily includes a spin-2 gravition (and spin-3/2 gravitino), a fundamental particle that carries the gravitational force. Finally, the prediction of superpartners creates viable candidates for dark matter: when R-parity is conserved, the lightest supersymmetric particle (LSP) is considered; when R-parity is violated, SM products of the LSP decay are studied instead. By permitting the LSP to have electric and color charges, RPV coupling introduces new BSM possibilities [9].

### 1.3 Recent Searches for BSM Predictions at the LHC

The ATLAS Collaboration searches for signatures of physics beyond the Standard Model via the Large Hadron Collider (LHC), discussed in Chapter 2. ATLAS employs theoretical methods to predict the contents and locations of targets in its searches. Previous R-party-violating (RPV) searches have focused on the minimal B-L extension of the MSSM, which allows for the further decay of the LSP [9]. In this model, interactions that do not conserve differences in baryon and lepton number have small couplings, preventing rapid proton decay [10].

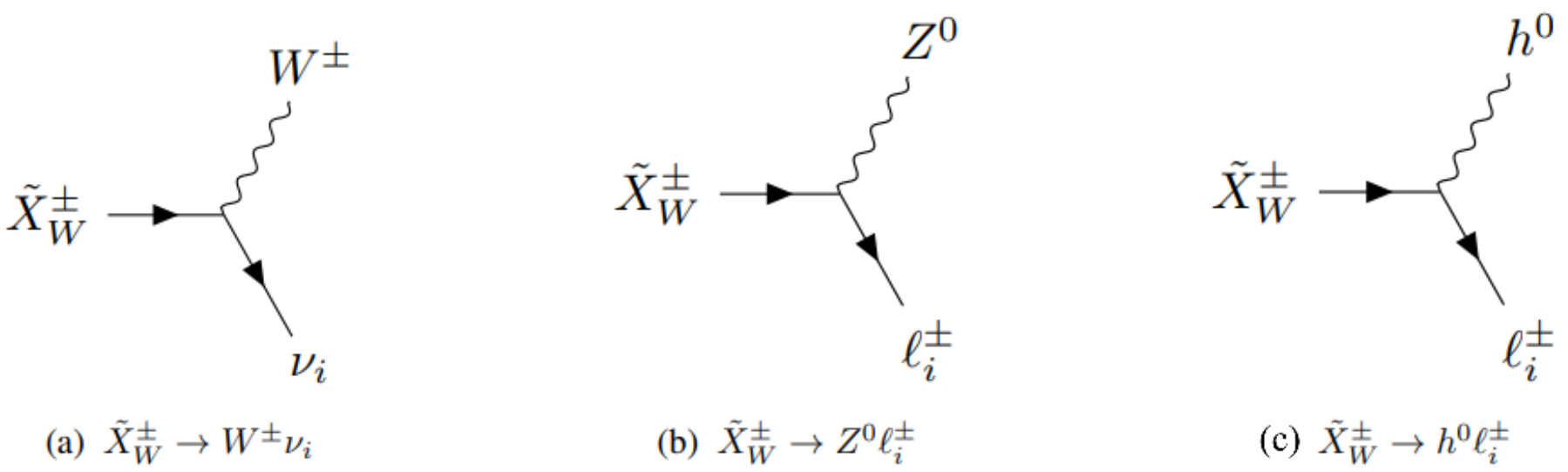

Figure 1: Possible decays of the Wino chargino LSP. The decay in Figure 1b) has been the focus of recent ATLAS searches.

Recent searches have focused on the decay of the Wino chargino LSP. Shown in Figure 1, the general massive chargino state $\tilde{X}_W^{\pm}$ decays into three potential RPV channels—each with final products that contain different levels of abundance and visibility in the detector. For example, the decay in Figure 1a can only be observed in the detector as missing energy, while the decay in Figure 1c yields quark and lepton traces that are difficult to interpret [9]. Most easily detected is the decay in Figure 1b, which produces many leptons from a single resonance. Recent pair production analyses have therefore focused on this decay [10].

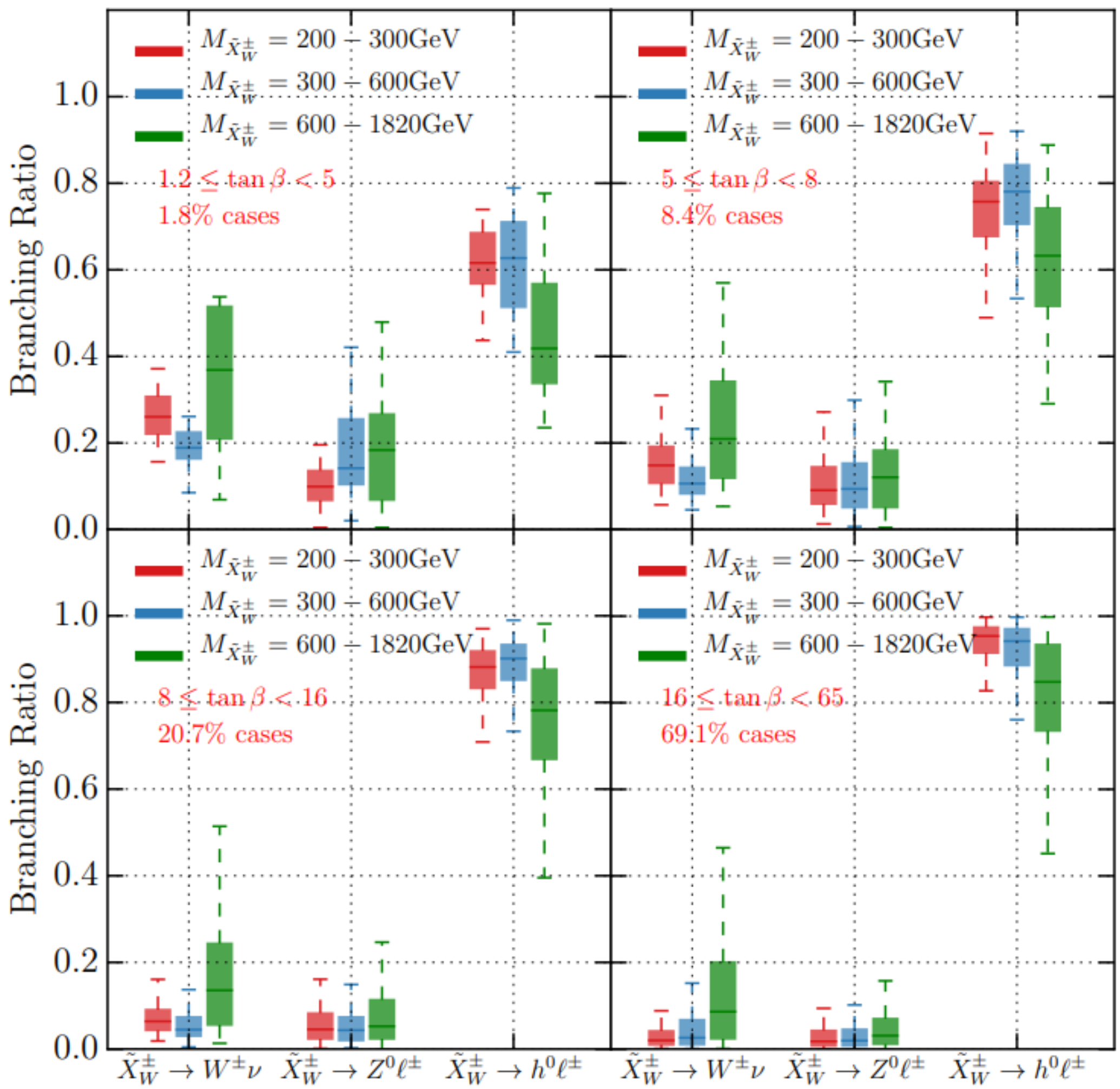

Figure 2: Branching ratios for different decay channels of the Wino chargino LSP. These ratios are depicted for a range of $\tilde{X}_W^\pm$ masses and tan β.

The branching ratios for the above channels have been calculated for a range of masses and tan β, using a normal neutrino hierarchy. According to Figure 2, $\tilde{X}_W^\pm \rightarrow h^0 \ell_i^\pm$ products are strongly favored for chargino masses of 200-1820 GeV: in fact, 200-300 GeV masses with tan β between 8 and 16 have median branching ratios of 0.064, 0.0445, and 0.882 for the a, b, and c channels, respectively. Though the $\tilde{X}_W^\pm \rightarrow h^0 \ell_i^\pm$ channel occurs most frequently, it has not yet been subject to an analysis.

### 1.4 Using Jets to Improve Future Search Sensitivity

Searching for the favored Wino chargino decay channel complements previous searches in its hunt for chargino pair production, whose final state consists of two leptons and two Higgs bosons.

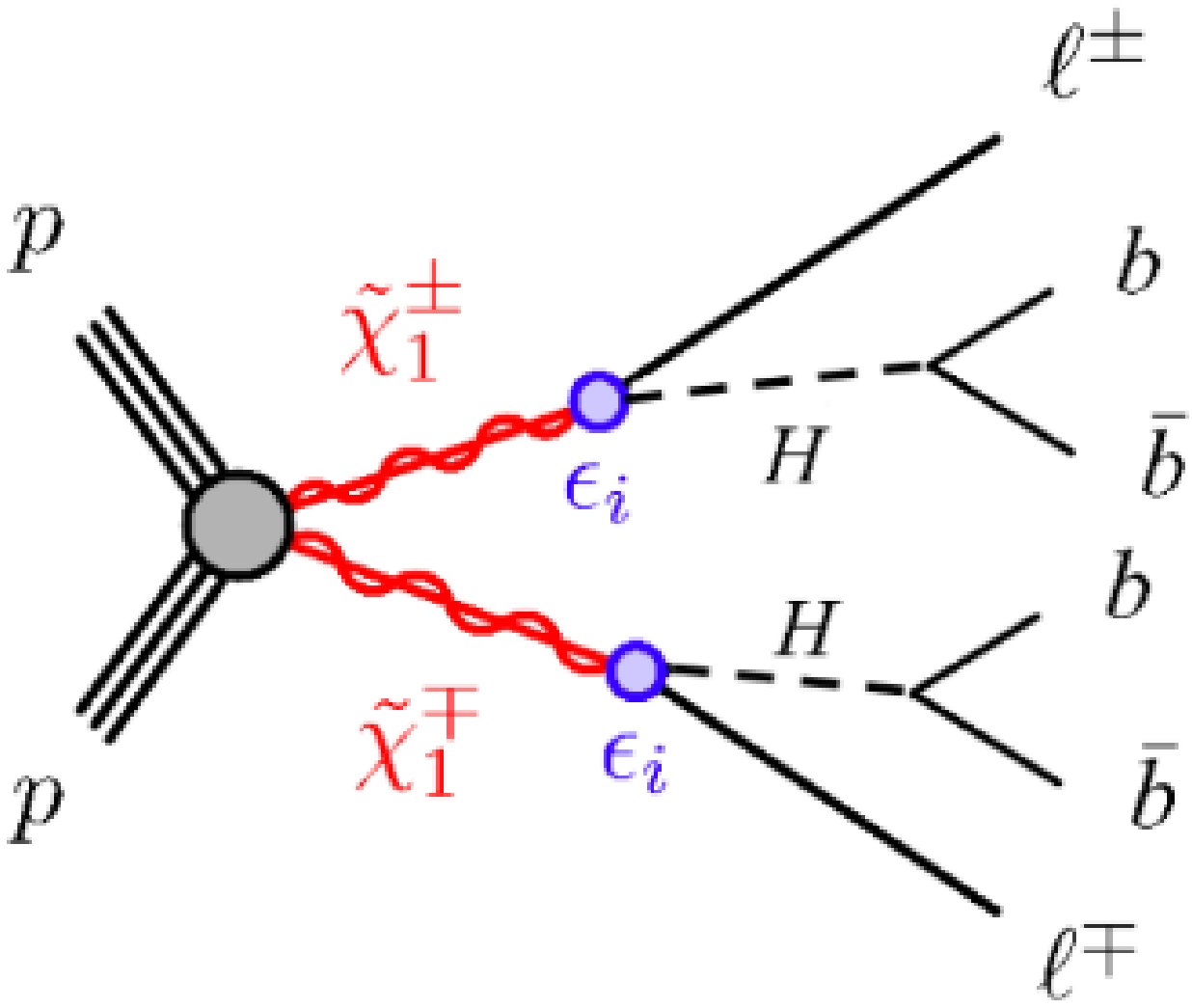

Figure 3: Feynman diagram of the favored Wino chargino LSP decay channel. Without the use of large-R jets, there are 6 possible b-quark pairings for Higgs reconstruction. When large-R jets are used, b-quarks are already paired for Higgs reconstruction.

These Higgs bosons decay 32.4% of the time to two pairs of b-quarks, whose high-momenta trajectories are captured with the use of jets [11]. By clustering showers of particles into calculable objects, jets reduce combinatorics in decays with difficult-to-pair products: large R (R = 1.0) jets combine their decay products into one jet that spans 2 radians, while small-R (R = 0.4) jets are used to further refine individual b-quark trajectories [14].

Large-R jets are particularly useful for simplifying the Higgs to b-quark decay, whose reconstruction would otherwise have 3 possible b-quark pairings. Large-R jets instead use the largest BR of Higgs to 2 b-jets in reconstructing each Higgs boson, allowing for a large jet mass and sufficient b-tagging information. These jets are most successful at reconstructing Higgs bosons with high transverse momenta (pT), where decay products are most collimated. Conversely, large-R jets are least successful at clustering decay products that are not highly collimated: because these jets attempt to capture 2 b-quarks within a set radius, they may incorrectly pair products whose distance apart is greater than the distance that the jet spans [4, 12].

Small-R jets capture individual b-quark trajectories, and they can be used to reveal specific information about decays. These jets are most successful at reconstructing decay products of lower pT Higgs bosons, as they can cluster b-quarks within larger distances of one another. These jets are least successful at capturing b-quark products of high pT Higgs bosons: because small-R jets do not cluster nearby products, the merging of jets that occurs in capturing highly collimated products results in objects that cannot be used [20].

Reconstruction techniques with collimated jets are crucial for improving the search's sensitivity. An emphasis on these high energy events reduces pileup, unrelated lower energy collisions that distract from the data of interest. Likewise, using such techniques improves mass resolution. Optimizing the use of jets is thus crucial in making precision measurements of Higgs production at different pTs [13-14].

This paper identifies parameters that improve jet performance by optimizing simulated reconstruction efficiency for different jet types. The paper is organized as follows: the ATLAS Detector is discussed in Chapter 2, simulated samples are discussed in Chapter 3, object reconstructiongg is discussed in Chapter 4, simulated results are discussed in Chapter 5, truth data analysis is discussed in Chapter 6, and conclusions are discussed in Chapter 7.

## 2 The ATLAS detector

The ATLAS detector is a general-purpose detector with a forward-backward symmetric cylindrical geometry and approximately 4 pi coverage in solid angle[1] [19]. It identifies BSM interactions by accelerating protons, whose head-on collisions form beams of new particles [15-17]. To measure different properties of these particles, it uses an inner tracking system, electromagnetic and hadronic calorimeters, and a muon spectrometer.

The Inner Detector reconstructs charged particle tracks in the pseudorapidity region $|\eta| < 2.5$. Its location within a 2 T axial magnetic field allows it to measure charged-particle momenta, which must initially be zero in the transverse direction [4, 10]. The Inner Detector measures radial tracks via silicon pixel, silicon microstrip, and transition radiation tracking detectors. Specifically, the transition radiation tracker (TRT) identifies electron trails, which ionize the xenon- or argon-based gas mixture within the device's many straws [16].

Located above the inner detector are the electromagnetic and hadronic calorimeters, which measure particle energy in the region covering $|\eta| < 4.9$ [10]. To collect energy, high density "absorbing" layers within the calorimeters induce particle showers, halting incoming particles. The energy of these showers is then measured by interspersed "sampling" layers [10, 15]. It is via large energy deposits in the hadronic calorimeter that b-quark-initiated jets are identified [4].

Because muons are too massive to deposit significant energy in the calorimeters, the muon spectrometer (MS) is used to measure muon momenta in the region $|\eta| < 2.7$. [10].

1 ATLAS uses a right -handed coordinate system. Pseudorapidity and angular distance are measures used to give information about the transverse plane, which is utilized because it is invariant to longitudinal boosts. Pseudorapidity is defined as $\eta = -\ln \tan(\theta/2)$, while angular distance is defined as $\Delta R = p\ (\Delta\eta)\ 2 + (\Delta\varphi)$ [21].

The MS's trigger and tracking chambers surround the calorimeters, measuring muon deflection in a toroidal magnetic field [15-17].

Finally, the ATLAS trigger system is used in reconstructing real and simulated data events. It is made up of hardware- and software-based triggers, which detect the locations and identities of particles [17].

## 3 Simulated Samples

Monte Carlo (MC) techniques were required to simulate interactions of decay products inside the detector. These techniques ultimately evaluate a search's expected sensitivity, tuning event selection and estimating predicted event yields and kinematic shapes [18].

Small- and large-R jets were externally constructed by ATLAS using the anti-kt algorithm, meeting its requirements for MC simulations [4, 5, 10]. Using electron, muon, and jet data, this paper was then able to reconstruct chargino information with ROOT software (See Chapter 4).

## 4 Object Reconstruction

Because this paper focuses on parameters that optimize jet accuracy, jets were first reconstructed using MC simulation. Unlike detector data, MC simulation gives much more detailed information about decay products that can then be used to explain findings from real data.

### 4.1 Lepton Selection

Events were tagged as EE, EM, ME, or MM, corresponding to events where both leptons are electrons, one lepton is an electron and the other is a muon, and both leptons are muons. For events with two electrons (muons), the electrons (muons) with the highest and second highest transverse momenta (pT) were chosen. For events with both electrons and muons, the electron and muon with the highest pT were identified. Their four-momenta were then calculated.

### 4.2 Large-R Reconstruction

For each event, large R jets were accepted if the number of large-R entries were at least 2. The four-momenta were then calculated for both jets. Because large-R jets envelop 2 b-quarks, Higgs boson information is already reconstructed.

To reconstruct the two chargino masses, the lepton and Higgs four momenta were added. Because there are two possible pairings of the two jets and two leptons, the momenta of each combination were calculated. These vectors were then manipulated to determine the invariant masses of each potential pairing.

To select which two combinations are correct, mass asymmetry was used. Mass asymmetry is defined as $A = \frac{|m_1 - m_2|}{m_1 + m_2}$, and it is used to determine the fraction difference between masses $m_1$ and $m_2$. Because the two chargino masses should be approximately equal, the pairs that minimize this value are selected. Selecting charginos with such methods not only increases accuracy but reduces contamination from background events [4]. Resulting chargino information was further kept if the mass asymmetry was less than 0.2.

### 4.3 Small-R Reconstruction

For each event, small-R jets were accepted if the number of small-R entries were at least 4. The four-momenta were then calculated for all jets, labeled jets 1 through 4 by leading pT.

Unlike large-R jets, small-R jets do not enclose two b-quarks that may then be used to reconstruct the Higgs bosons. Instead, the two correct reconstructions are selected from 3 possible pairings of small-R b-jets. This is done by finding the two pairs that have the smallest combined mass asymmetry between themselves and the 125 GeV Higgs bosons. After this reconstruction, the same procedure is used as that of the large-R jets in pairing the Higgs with the correct lepton to obtain chargino information. Chargino information was only kept if the mass asymmetry was less than 0.2.

### 4.4 "Varied-R" Reconstruction

Most events were captured by both small- and large-R jets, which were used to derive chargino information. To test which parameters were most optimal for each type of jet, an algorithm was created to choose the jet type that most correctly reconstructed both Higgs masses per event. Further information about the most successful jets could then be derived from "varied-R" jet data.

To determine which jets were most successful, the mass asymmetries between the small-R-reconstructed Higgs bosons and the 125 GeV Higgs masses were first calculated and combined. The mass asymmetries between the large-R-reconstructed Higgs bosons and the 125 GeV Higgs masses were also calculated and combined. These combined asymmetries were then compared, and the jets with the lower value were chosen. If one of the jet types did not have enough entries per event to be originally selected, the reconstructed Higgs bosons from the other jet type were observed.

Chargino information was kept if both reconstructed Higgs masses were within 20 GeV of the true Higgs mass. The mass asymmetry between the two reconstructed Higgs was also required to be less than 0.2, consistent with the original selection process for small- and large-R jets.

## 5 Simulated Results

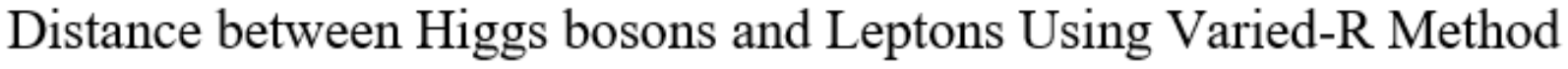

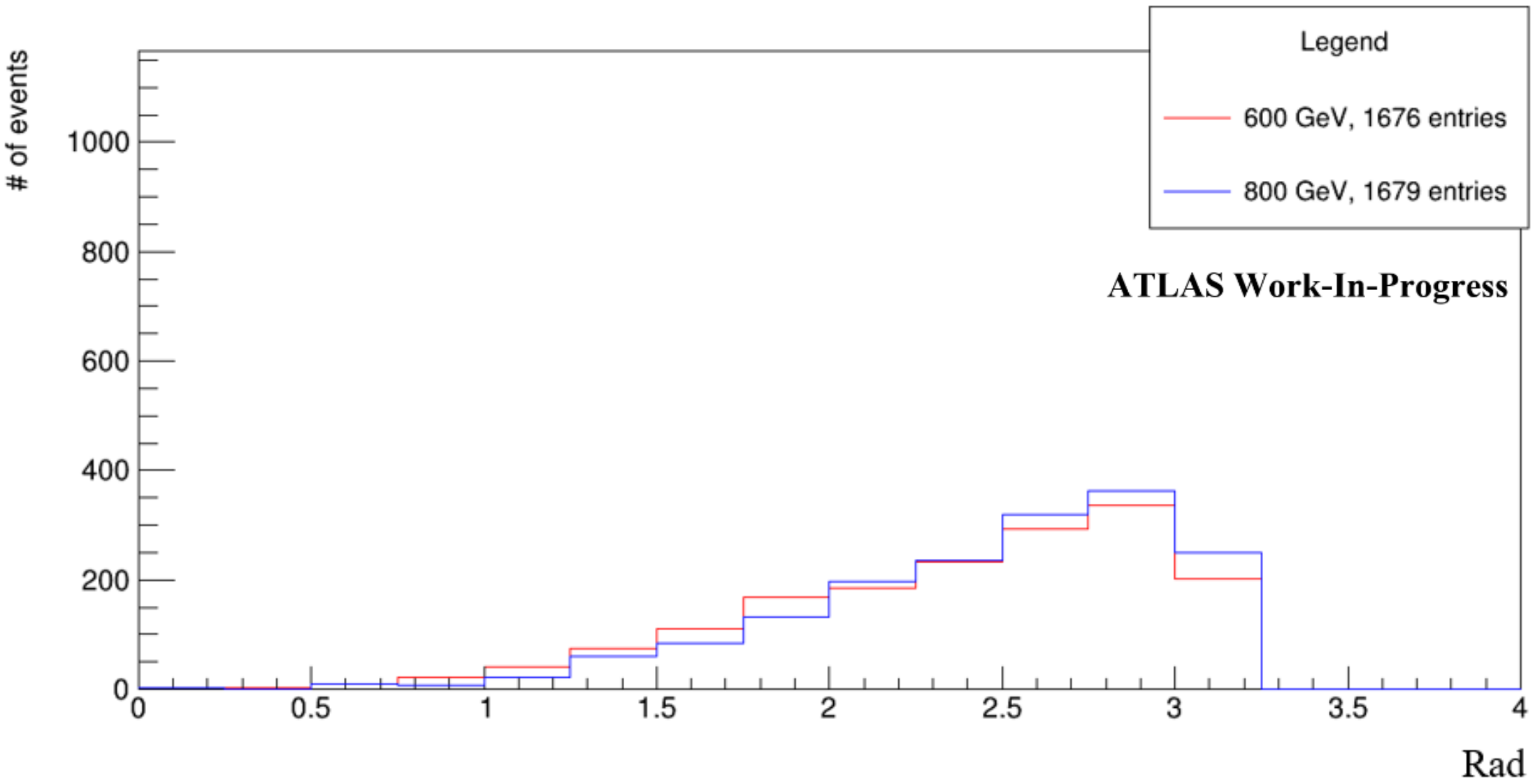

Figure 5: Distance between Higgs bosons and leptons for the most accurate jet types. This varied-R reconstructed data was plotted from both 600 and 800 GeV chargino information.

600 and 800 GeV chargino data were analyzed. When varied-R jet pairs were analyzed, it was found that the most accurate ones best agreed with Standard Model predictions of jet behavior. Figure 5 shows the distance between varied-R reconstructed Higgs bosons and their corresponding leptons. As chargino mass increased, this value increased in the lab frame. These findings corroborate theoretical findings, which predict that increases in chargino pT cause jet behavior to agree most with the back-to-back decay that occurs in the rest frame.

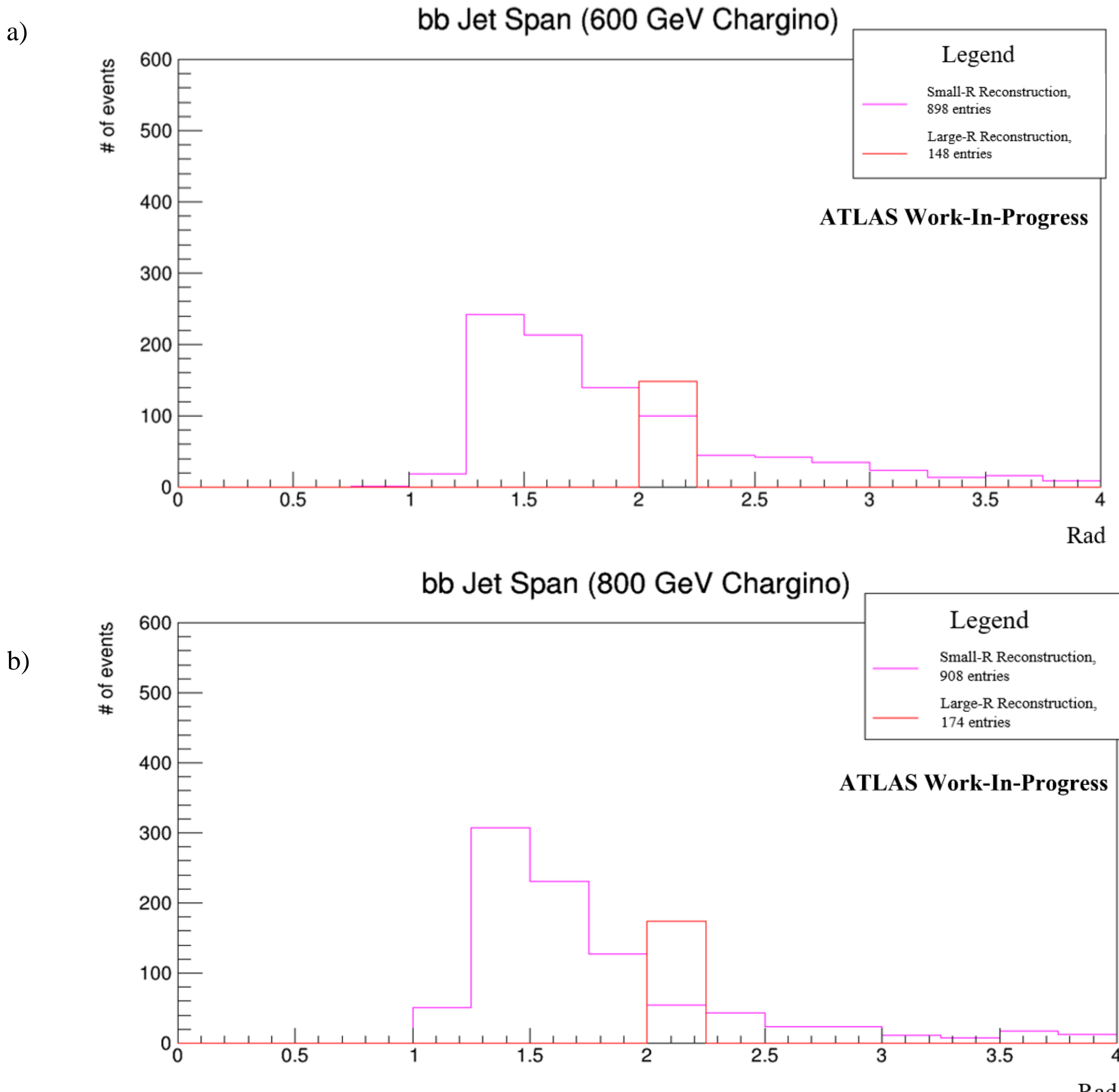

Figure 6: Jet span of the most accurate small- and large-R jets using a) 600 GeV chargino data and b) 800 GeV chargino data. These jets make up the varied-R data set.

Figures 6a and 6b show the jet span of the most accurate jets used for 600 and 800 GeV chargino reconstructions, respectively. These values decreased with increased chargino mass, agreeing with SM predictions that jet products become more collimated with increased pT.

These figures also reveal information about the accuracy of the jets that are used for certain jet spans: because large-R jets have a set jet span of 2 radians, any jet span beyond this value is an indication that small-R jets had to be used to accurately capture the event. 8.1% of the most accurate 600 GeV events and 5.6% of the most accurate 800 GeV events were not in the span of large-R jets. This agrees with SM predictions, in which large-R accuracy is expected to rise with pT.

Despite discrepancies in the ability of each jet type to capture events, most of the values in the figures were captured by the range of both jets. Such findings indicate that other conditions are relevant in optimizing which jet type is most accurate, especially for higher mass charginos.

**5.1 Higgs and Chargino Reconstruction**

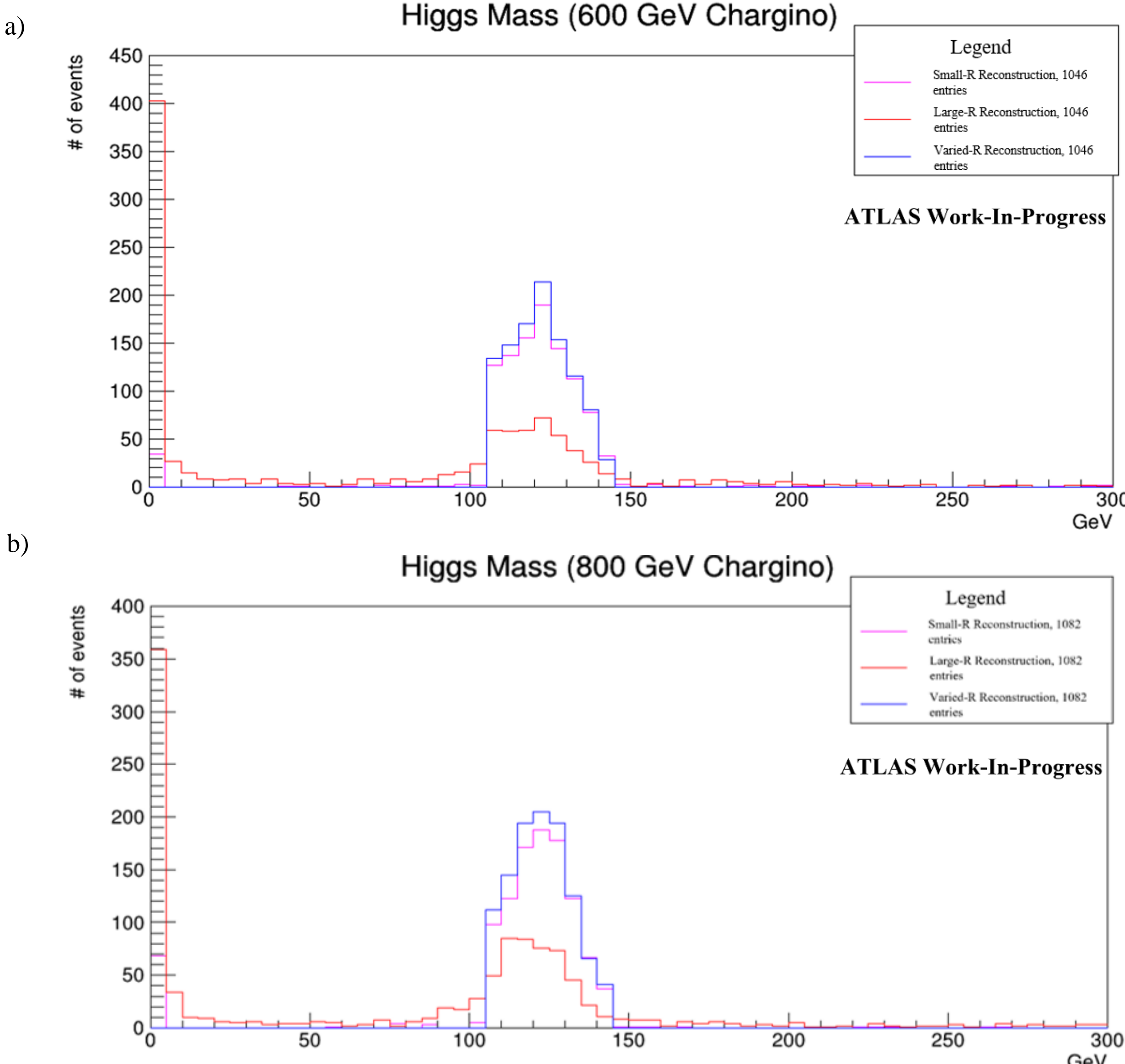

Figure 7: Reconstruction of Higgs masses with small-, large-, and varied-R jet types. a) shows reconstruction from 600 GeV chargino data while b) shows reconstruction from 800 GeV chargino data.

Figures 7a and 7b depict Higgs boson reconstruction for 600 and 800 GeV chargino masses, respectively. Both small-R and large-R reconstructions are shown, and the varied-R data plots the jet-type reconstruction with the more accurate Higgs masses. For events that capture less than two large-R jets, both Higgs masses were set to zero, indicating that the full reconstruction was unachievable. The same procedure is used for events with fewer than four small-R jets.

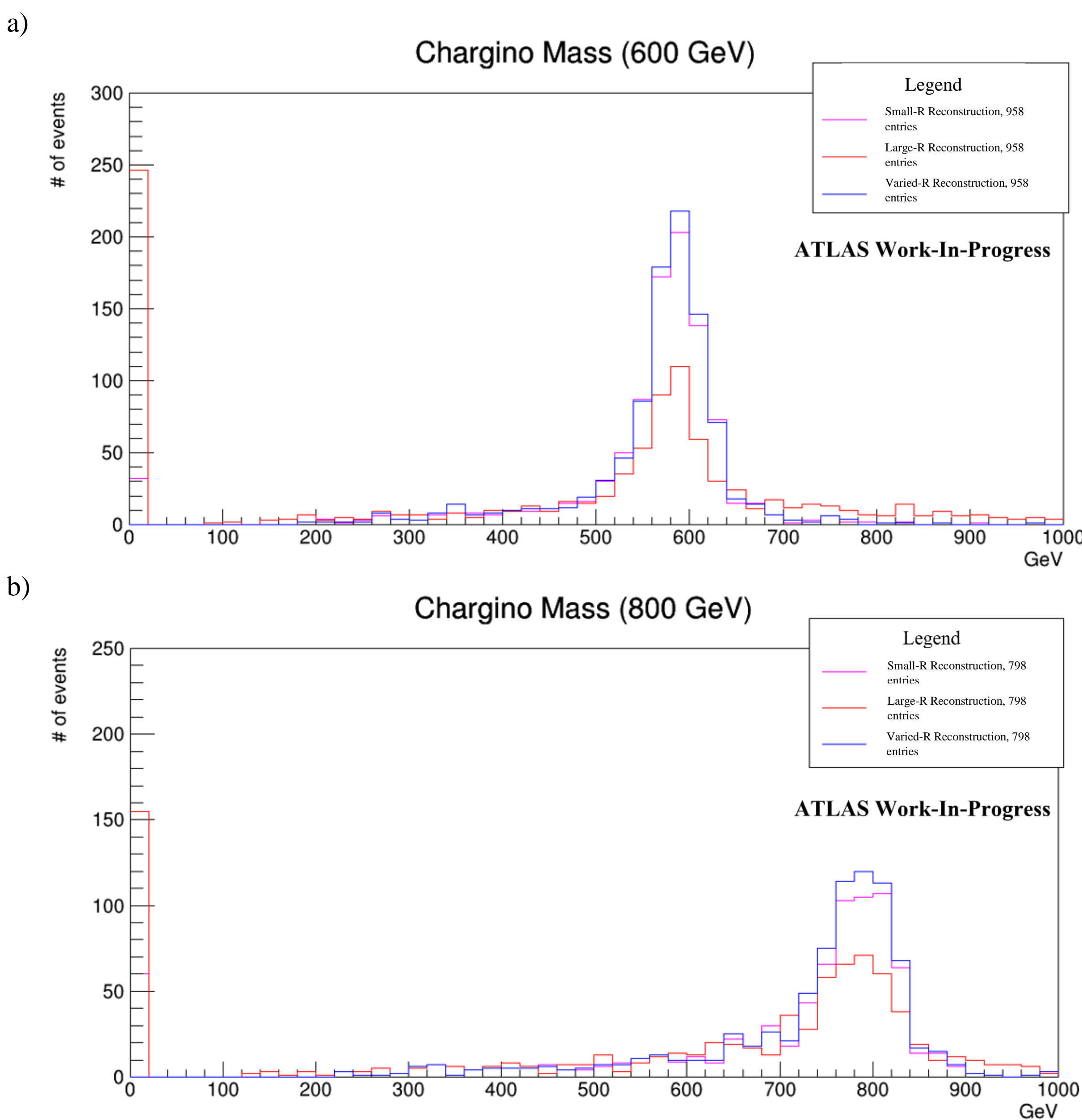

Figure 8: Reconstruction of chargino masses with small-, large-, and varied-R jet types. a) shows reconstruction from 600 GeV chargino data while b) shows reconstruction from 800 GeV chargino data.

The combined jet type method best rebuilt both Higgs bosons to an accuracy within 20 GeV of the true Higgs mass. This method not only optimized the use of jets near the Higgs mass but introduced data when one jet type could not capture an event. Because of this, both chargino masses were reconstructed more accurately with the varied-R method, shown in Figures 8a and 8b.

### 5.2 Discrepancies in Jet Performance

Small-R jets captured most of the events for the chargino masses more accurately than their large-R counterparts. They were most successful at reconstructing Higgs data from lower mass charginos: compared to large-R jets, they captured more events accurately with 600 GeV chargino data, and fewer events were rejected for insufficient information.

Large-R jets were not nearly as successful as their small-R counterparts at accurately reconstructing the Higgs boson for both chargino masses. Even so, the accuracy of this jet type increased with increased chargino mass: compared to large-R reconstruction with 600 GeV chargino data, large-R reconstruction with 800 GeV chargino information captured more events accurately and rejected fewer events for insufficient information.

Some discrepancies were seen between accepted jet types and their rejected counterparts, especially for lower chargino masses. These values were plotted against each other for 600 GeV charginos for further analysis. Figure 9 plots the distance between

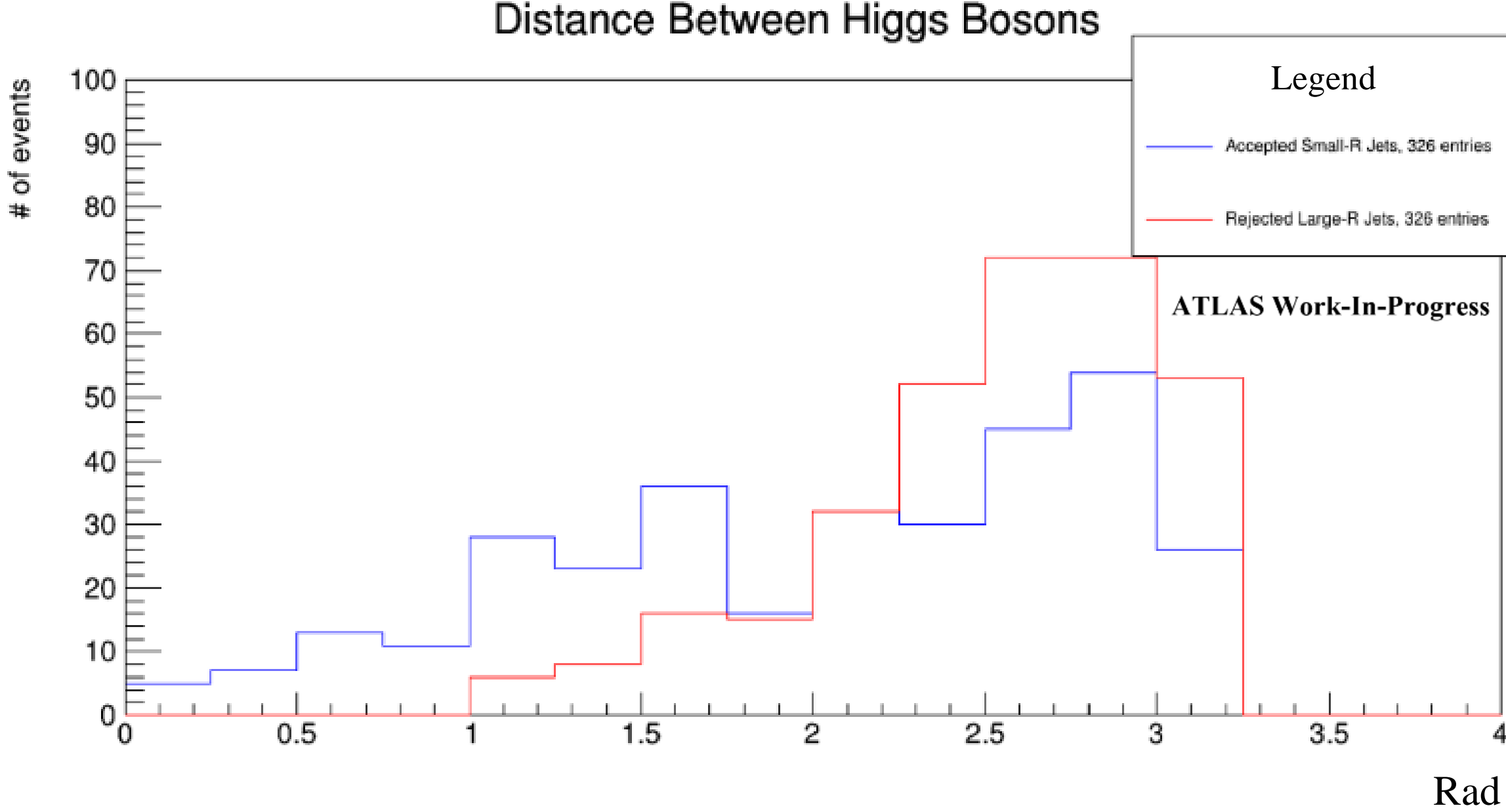

Figure 9: The distance between Higgs bosons, plotted for more accurate small-R jets and their rejected large-R counterparts. This is used to reveal event characteristics that play a role in selecting accurate jet types.

Higgs bosons (Higgs delta R) for accepted small-R jets and their rejected large-R counterparts: this plot indicates that small-R jets are optimal for reconstructing events whose Higgs bosons are both within the range of the large-R jet span. For these events, small-R jets correctly reconstruct the Higgs masses, which large-R jets do not have the ability to accurately cluster. Because the same discrepancies were not seen for accepted large-R jets and their rejected small-R counterparts, this plot indicates that the distance between Higgs bosons is a distinct event characteristic that can be used to select jet types for signal accuracy.

Some parameters were unclear in their relevance for selecting events. Event characteristics such as the distance between Higgs bosons and their corresponding leptons did not give significant discrepancies in jet type performance, and metrics such as the distance between leptons did not serve as an adequate predictor for values like jet span and Higgs radius. Likewise, while jets with a pT of approximately half the chargino mass were more often selected, it is unclear the extent to which jet type plays a role.

While some of these parameters were not highly relevant in selecting accurate jets, it is unclear whether a combination of these characteristics is relevant in selection. Because predicting these indicators proves useful for reducing the complexity of reconstruction algorithms, further exploring such combinations with neural networks is promising for improving jet selection.

## 6 Truth Sample Analysis

Truth samples were obtained for chargino masses of 150 GeV to 1400 GeV. Information about the correct chargino decay products, pairings, and b-quark-matched jets was then used to evaluate the efficiency of various reconstruction methods.

### 6.1 Truth Jet Performance

Simulated sample results suggest that preventing small-R jet merging improves reconstruction accuracy: this jet overlap was prevented in reconstructing truth data by ensuring that both the distance between b-quarks and the distance between Higgs bosons were at least 0.4 radians. Likewise, simulated sample results suggest that ensuring that large-R jets capture the correct number of products prevents inaccurate reconstruction: this was accomplished in truth data reconstruction by ensuring that the b-quarks were within the large-R jet span while both Higgs bosons were not. These parameters were used on truth data to determine the maximum accuracy of small- and large-R jet reconstruction across chargino masses. Figure 10 depicts this on the following page.

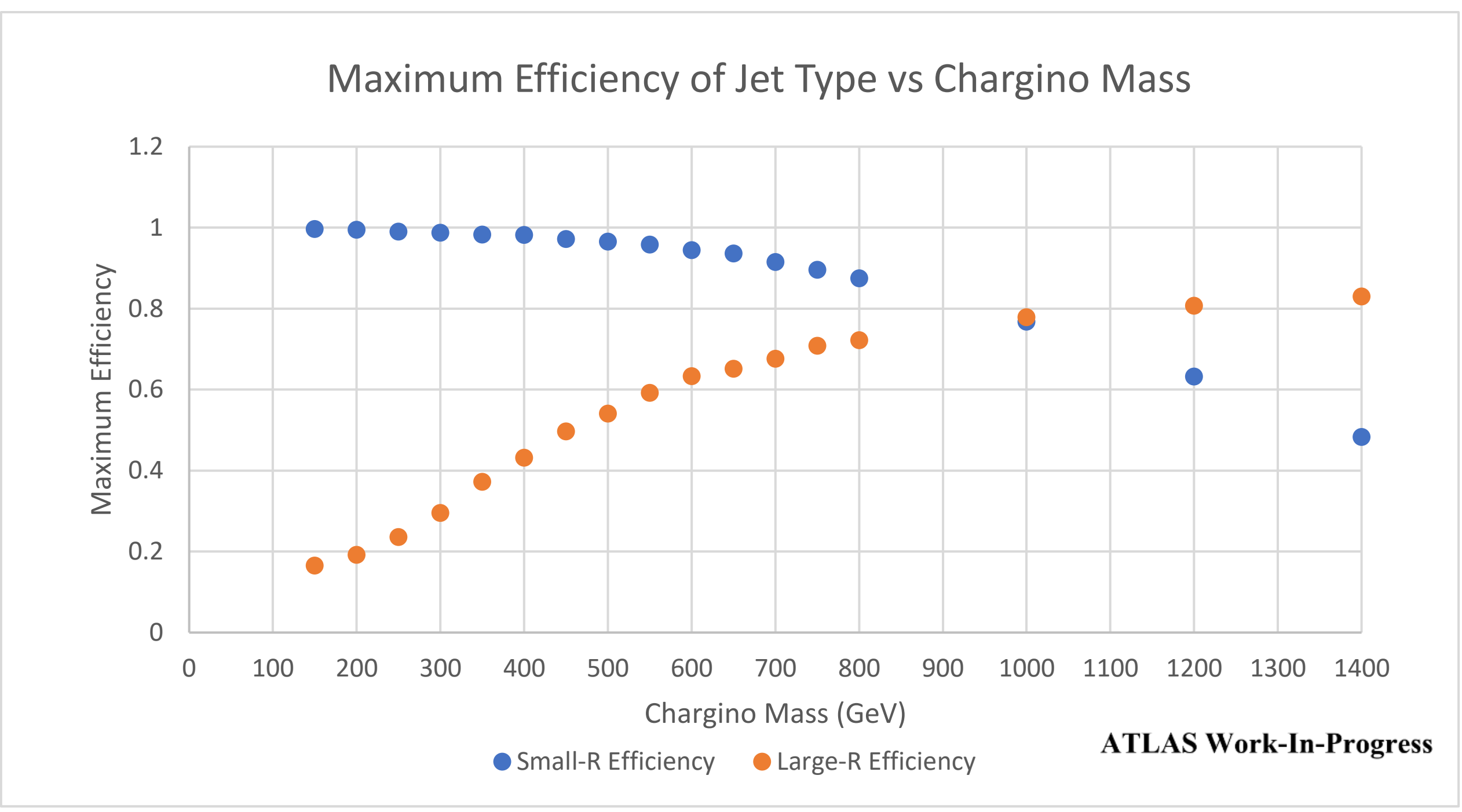

Figure 10: Maximum small- and large-R jet efficiencies across chargino masses. These efficiencies are defined by the ability of each jet to capture the necessary number of b-quarks and Higgs bosons.

Because of its higher maximum efficiency across most relevant chargino masses, small-R reconstruction was selected for further analysis.

### 6.2 Increasing Reconstruction Efficiency

Reconstruction efficiency was increased by focusing on small-R jet selection across chargino masses. Higgs boson and chargino masses were first reconstructed without preventing small-R jet merging. Figure 11 reconstructs Higgs masses from 600, 800, and 1000 GeV charginos, which were chosen because of their highly collimated products. While the proper 125 GeV Higgs mass was reconstructed, a nonnegligible number of small-R reconstructions peaked at approximately 250 GeV. These incorrect reconstructions increased with chargino mass.

Even when various common pairing methods were used, incorrect 250 GeV Higgs reconstructions remained. These pairing methods included minimizing the mass asymmetry and mass difference of small-R jet pairs, as well as reducing the distance between individual jets.

a)

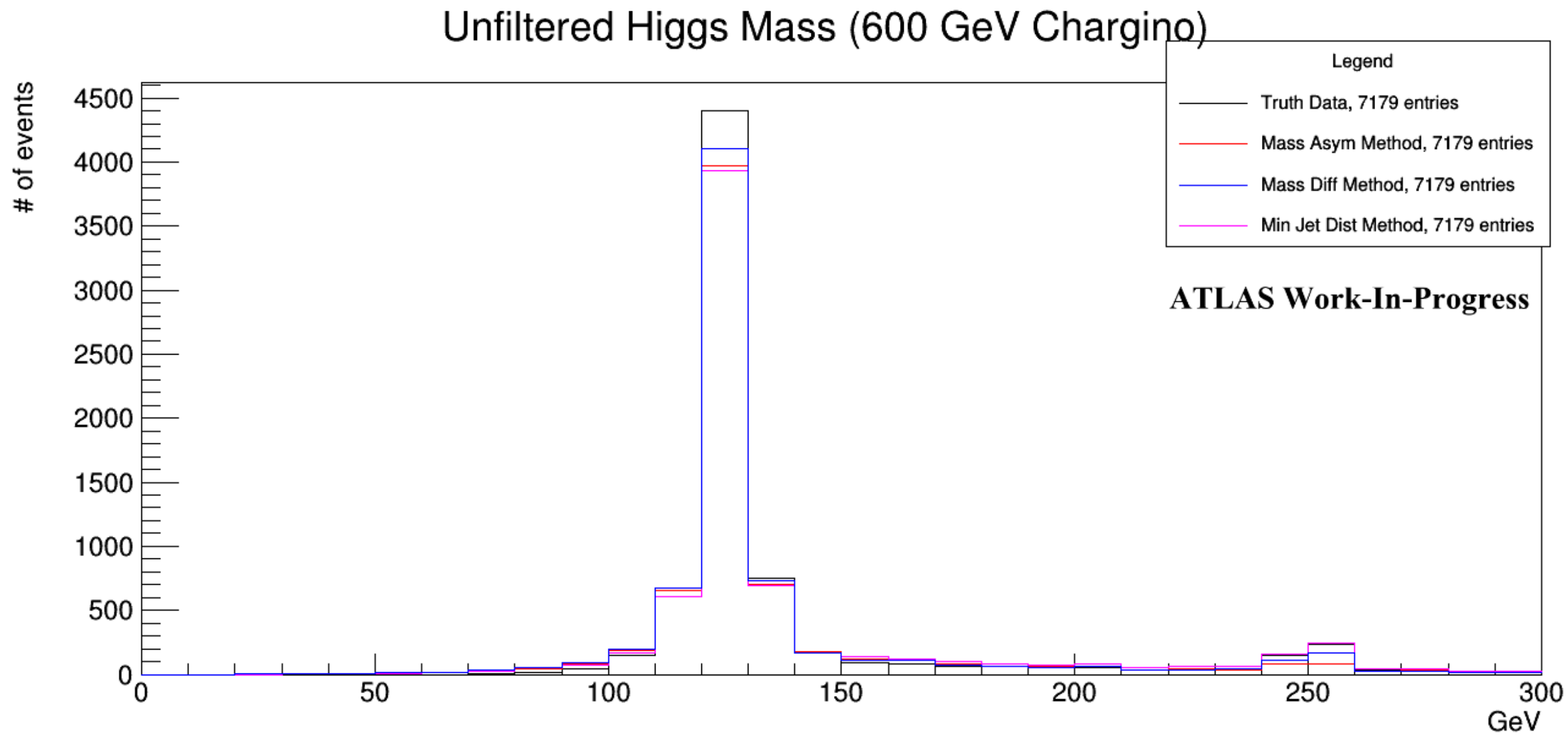

b)

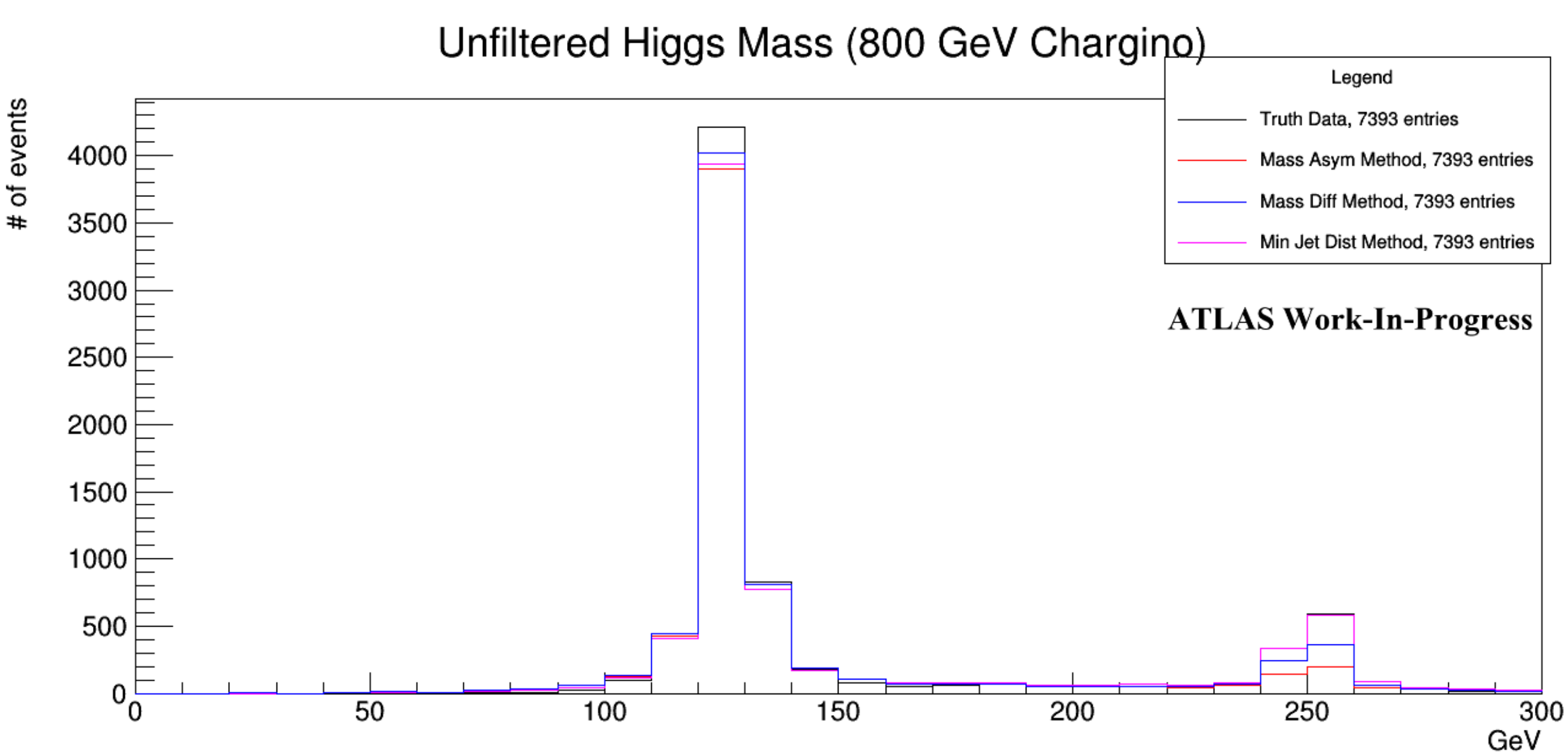

c)

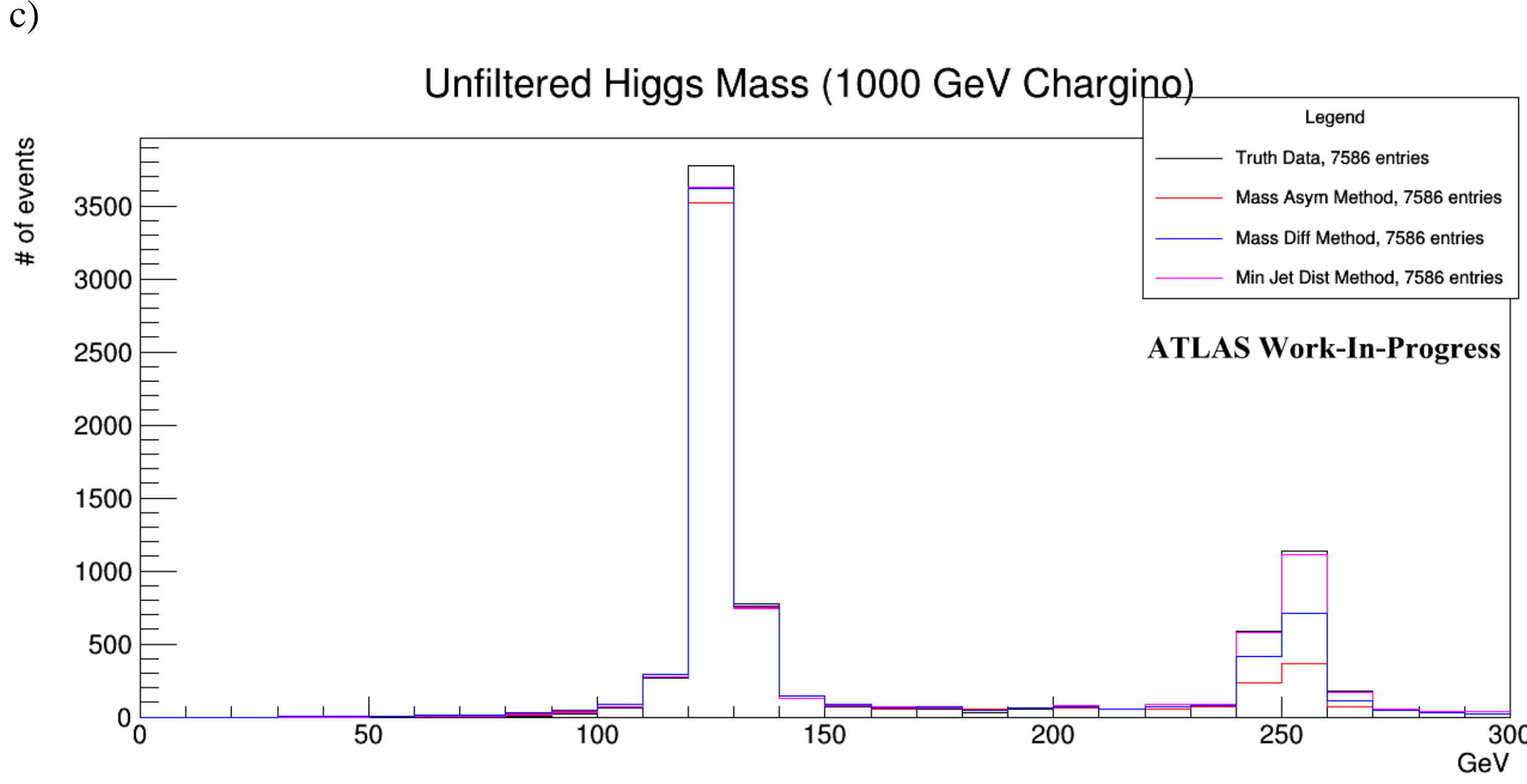

Figure 11: Reconstructed Higgs masses from a) 600, b) 800, and c) 1000 GeV chargino masses. This data did not filter decays that small-R jets inaccurately capture.

Likewise, further reconstruction of the chargino masses created an additional incorrect peak, especially at higher masses. These unfiltered reconstructions are shown in Figure 12.

a)

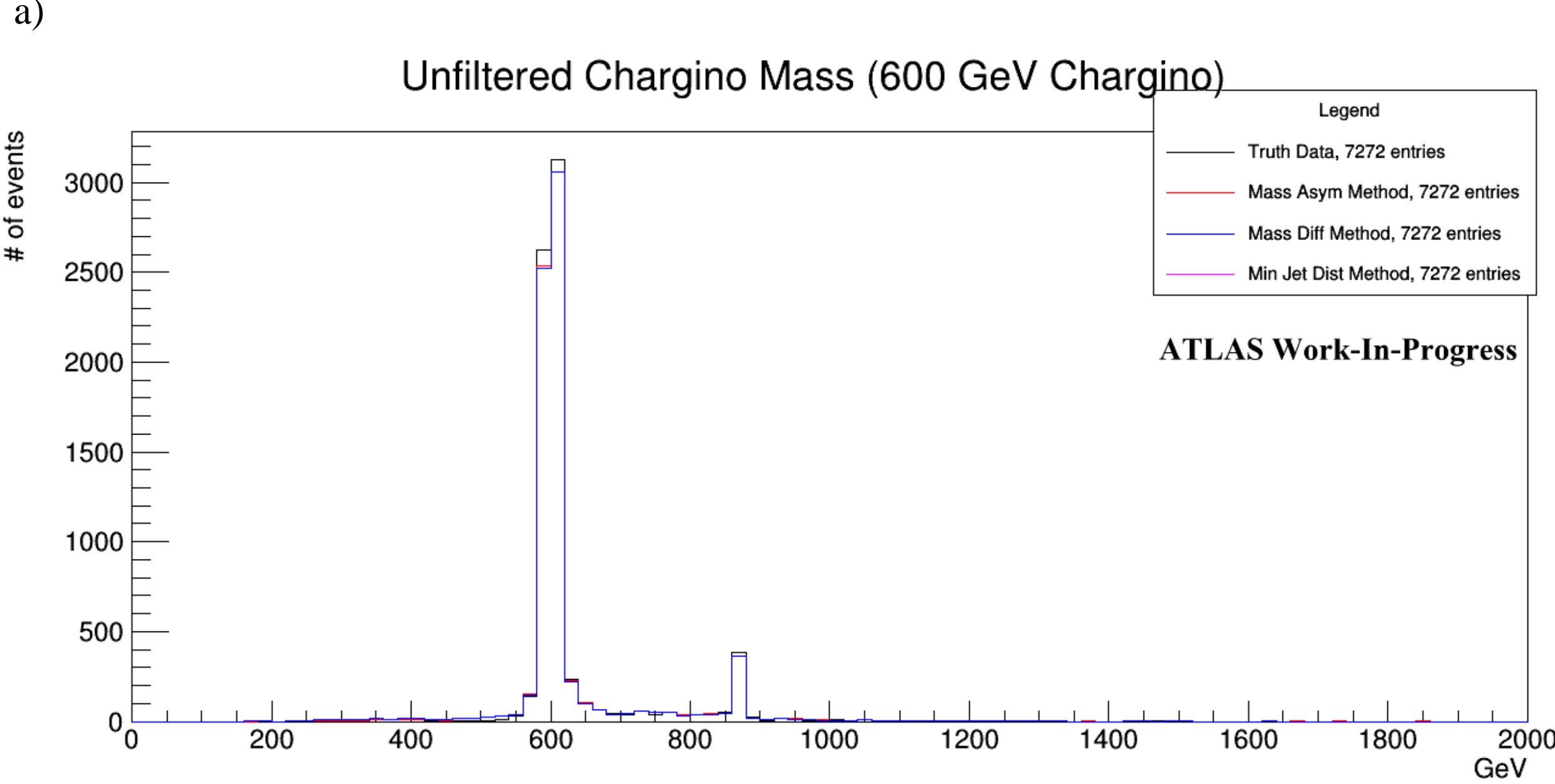

b)

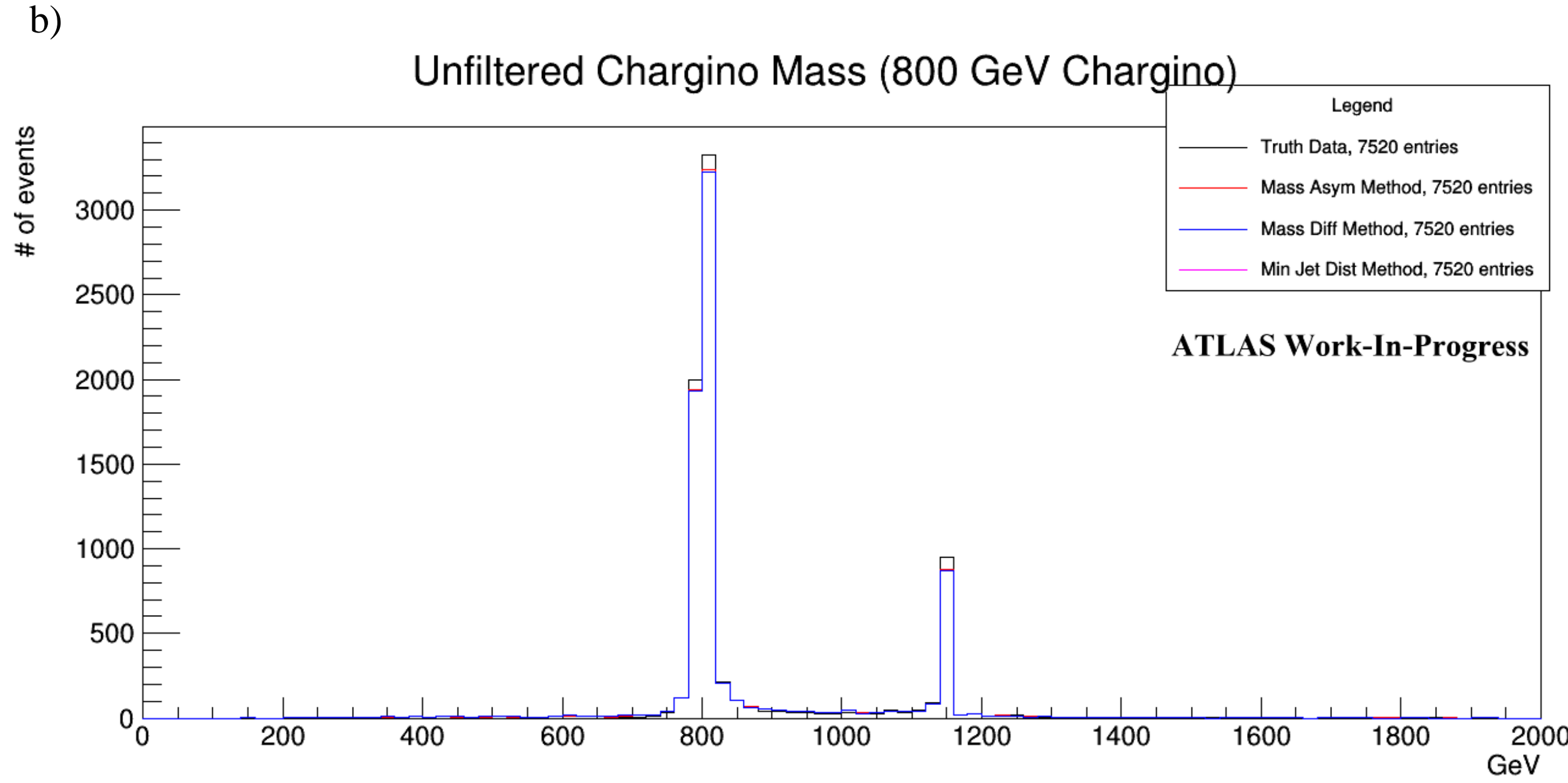

c)

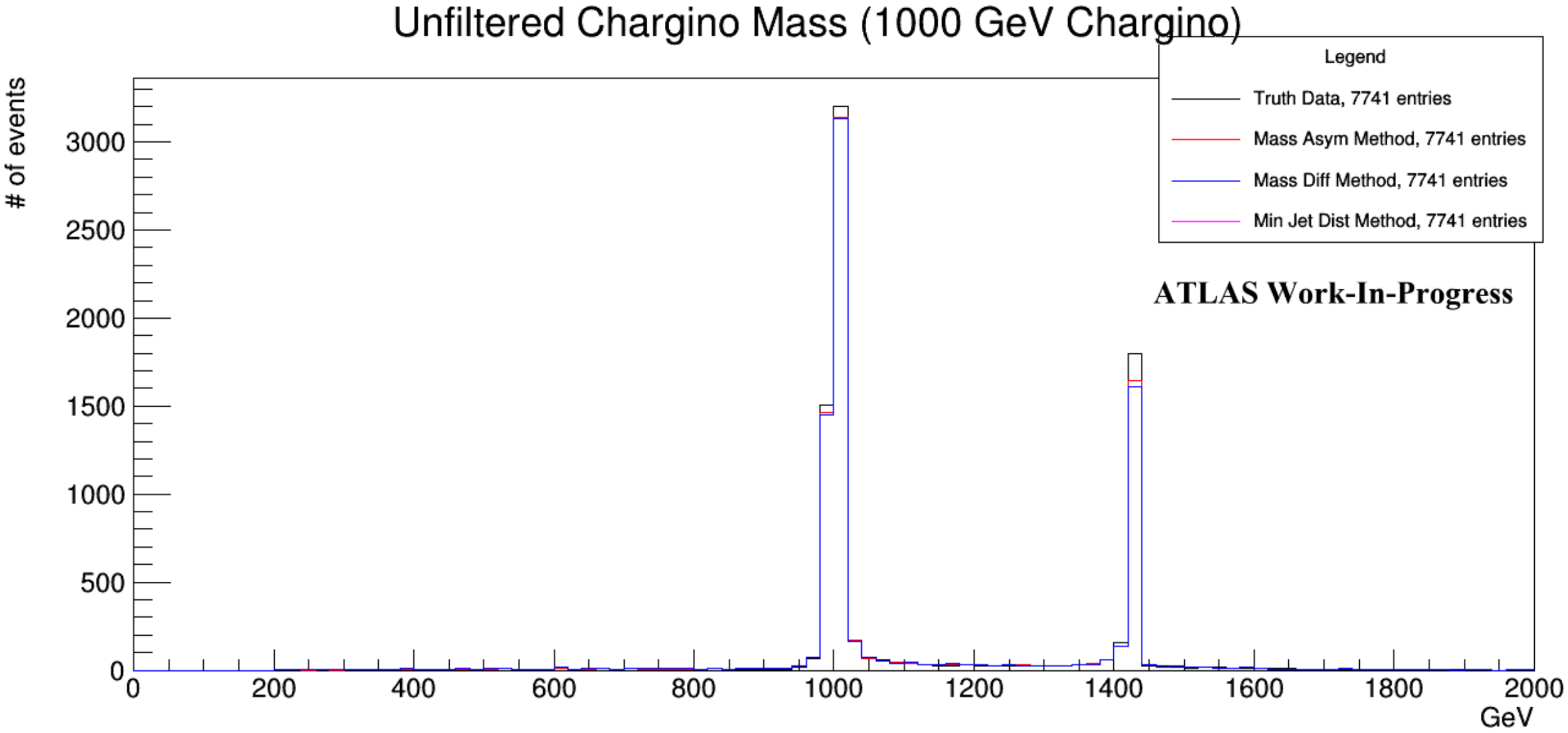

Figure 12: Reconstructed a) 600, b) 800, and c) 1000 GeV chargino masses. This data did not filter decays that small-R jets inaccurately capture.

These masses were then reconstructed after filtering out jet pairs and Higgs bosons whose respective distances were less than the small-R jet radius.

a)

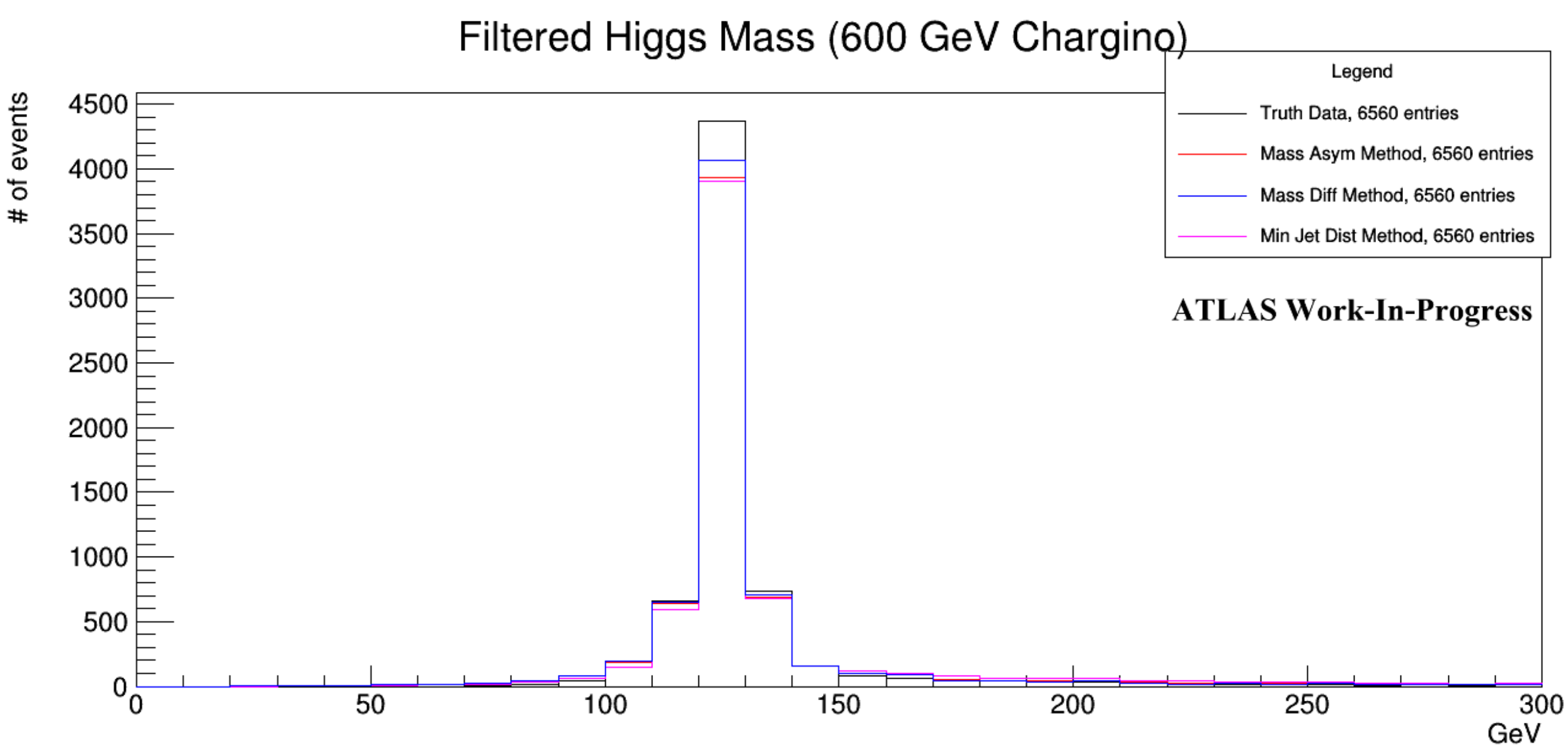

b)

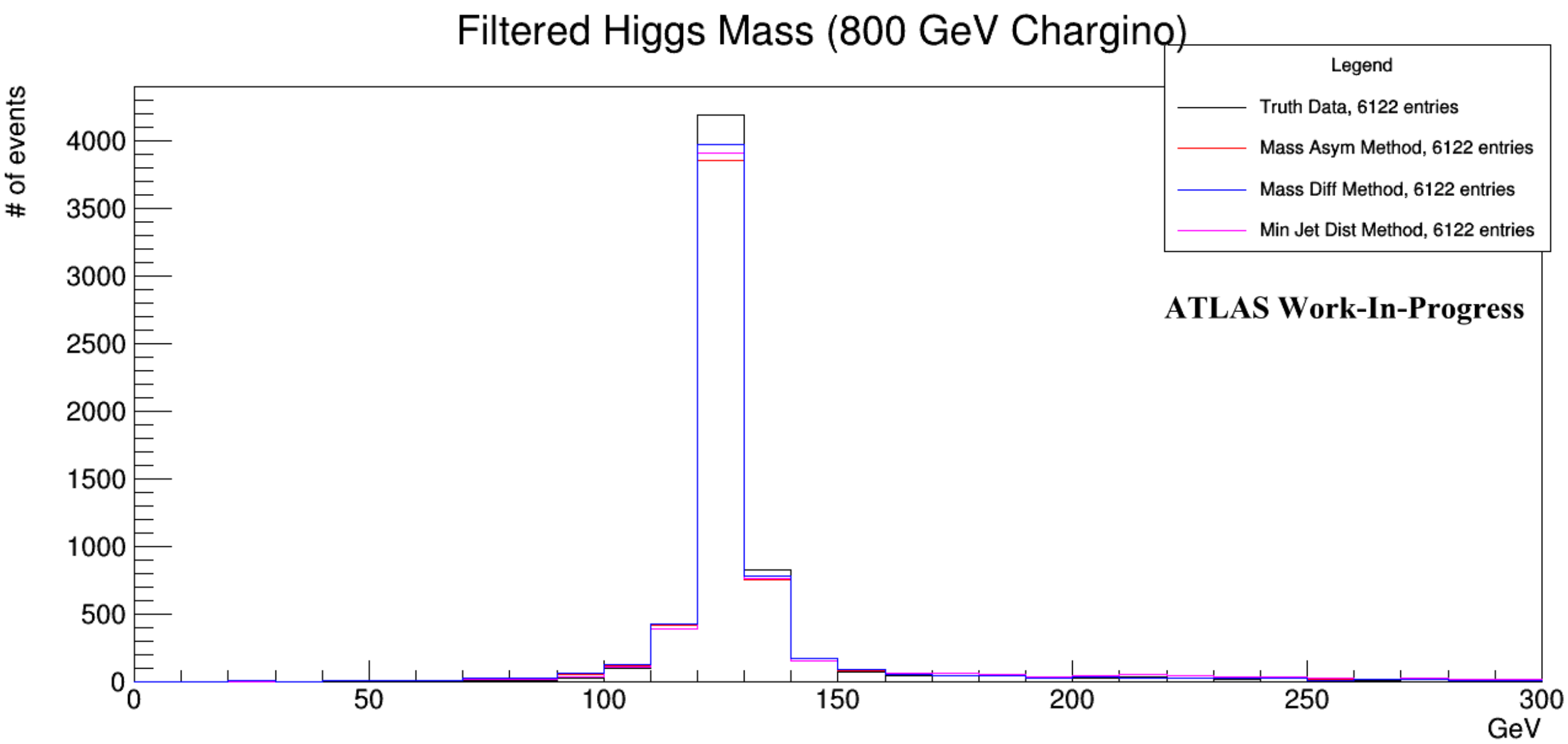

c)

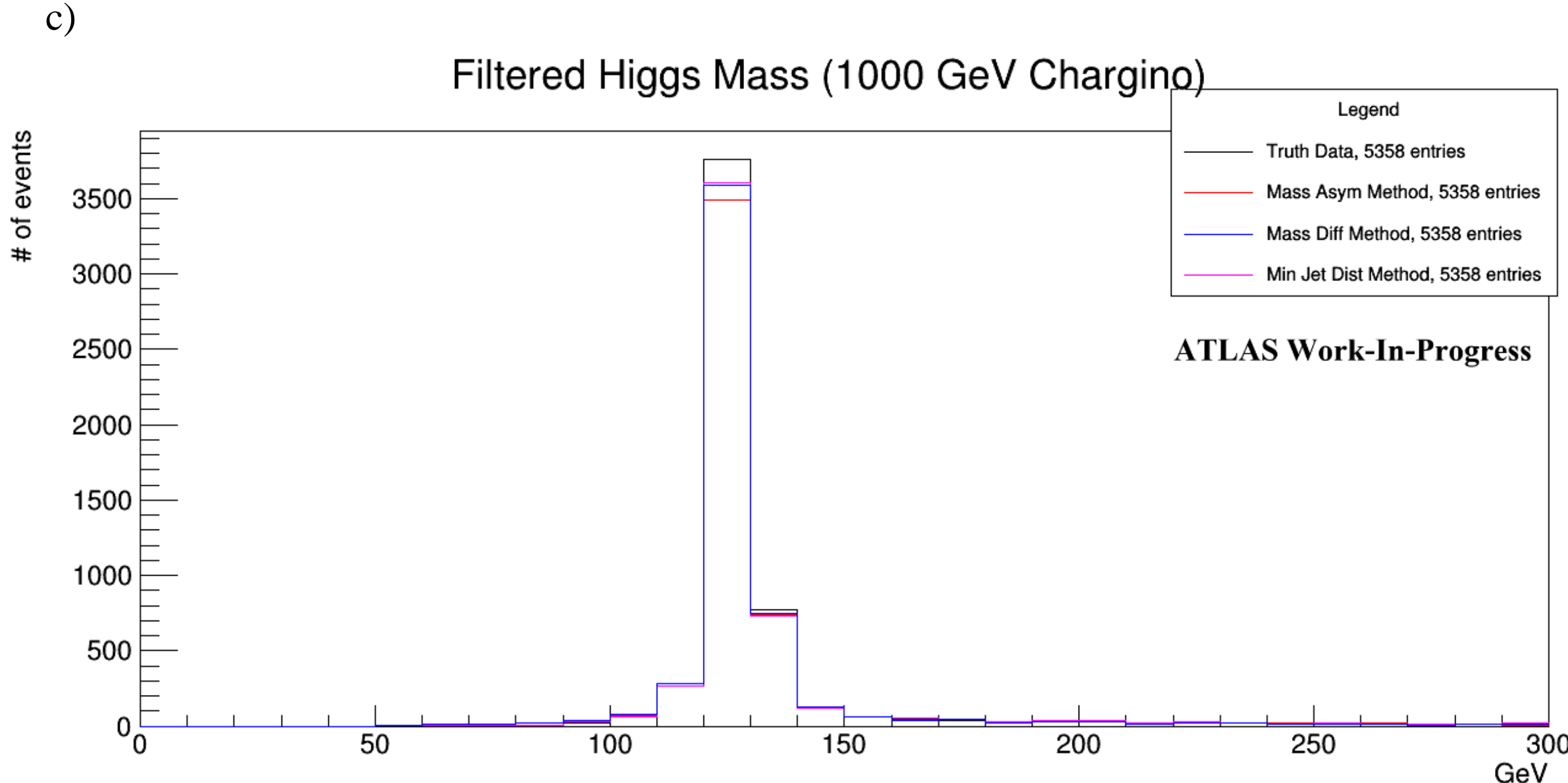

Figure 13: Reconstructed Higgs mass from a) 600, b) 800, and c) 1000 GeV chargino masses. This data filtered decays that small-R jets inaccurately capture.

It is clear that filtering decays by jet and Higgs distance impacted the accuracy of the reconstructed Higgs mass. Employing this algorithm also did not significantly reduce the total number of events captured, which proves useful for larger analyses.

Finally, a similar process was used to filter chargino reconstruction for increased small-R jet accuracy, shown in Figure 14.

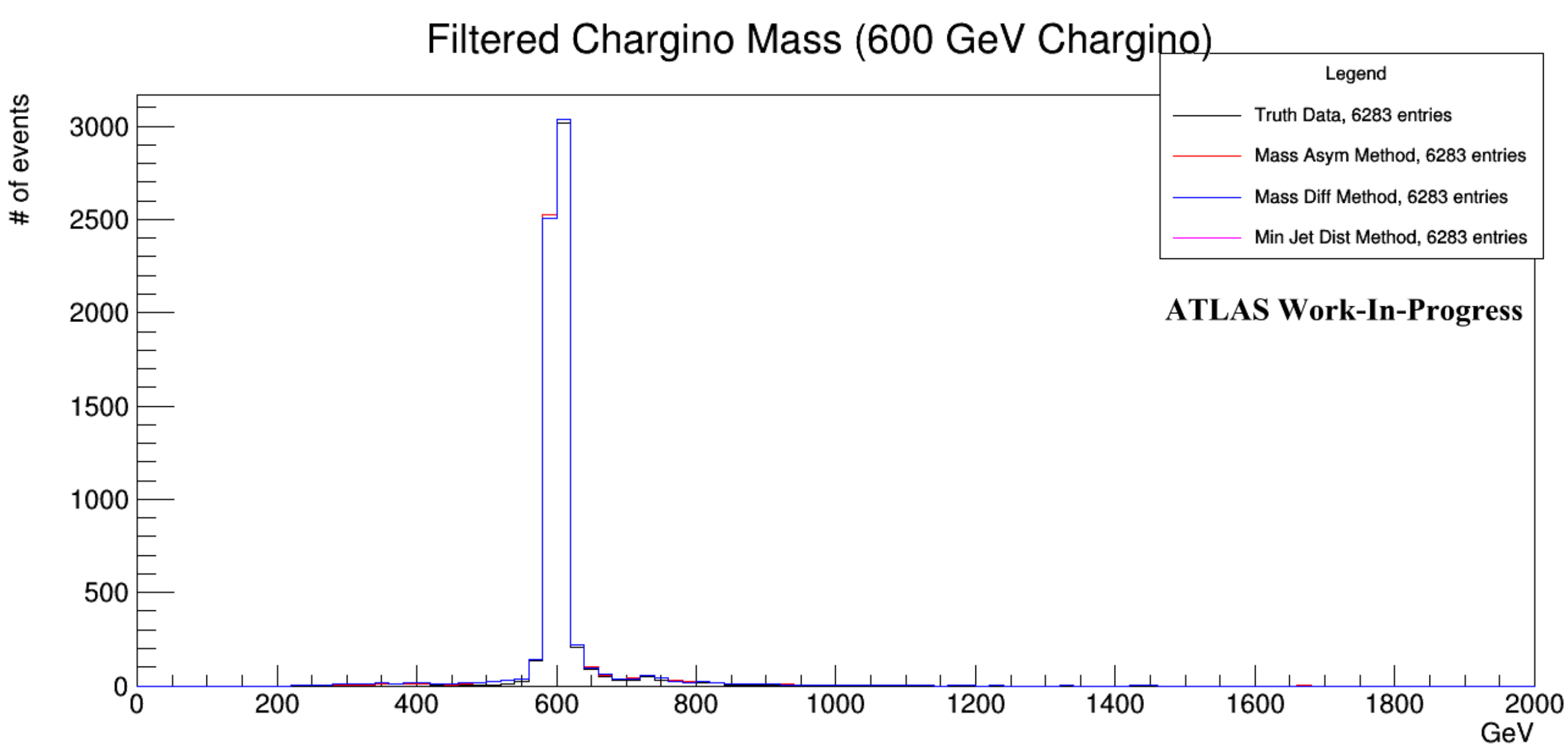

b)

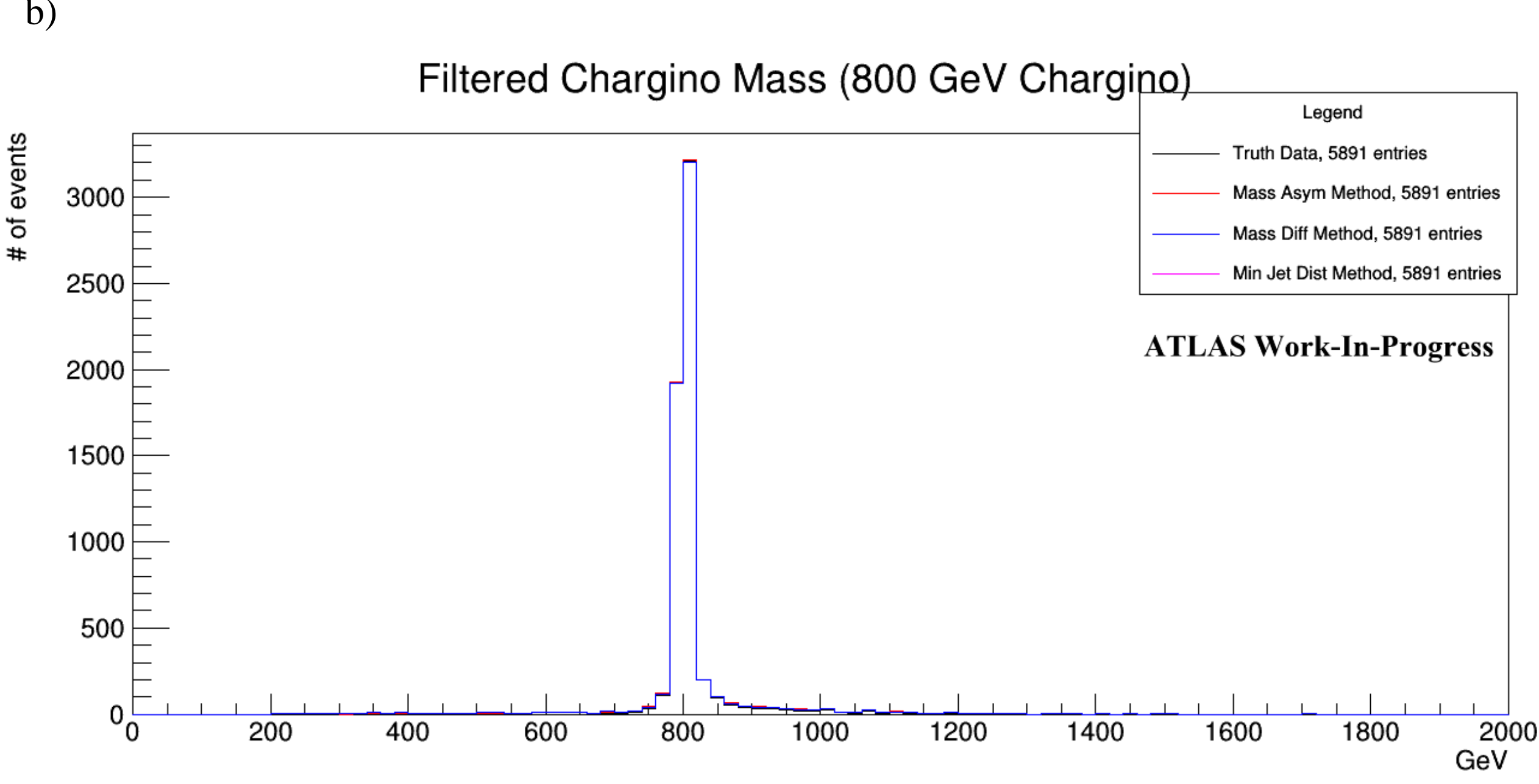

c)

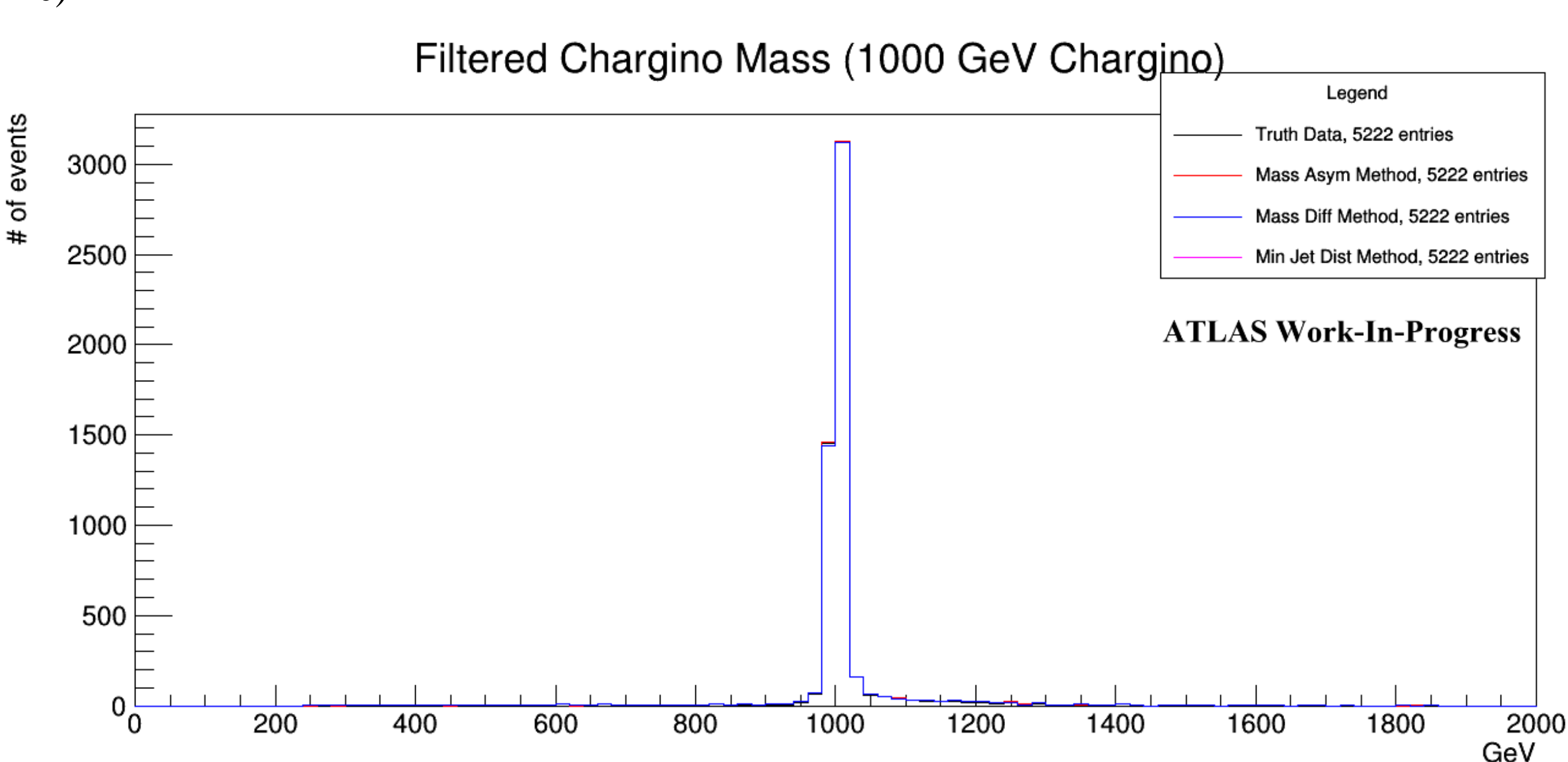

Figure 14: Reconstructed a) 600, b) 800, and c) 1000 GeV chargino masses. This data filtered decays that small-R jets inaccurately capture.

The improved reconstruction methods above notably increased efficiency without significantly reducing the total number of events analyzed. What was most relevant to this improvement was not the various small-R pairing methods used but the process by which reconstructed events were filtered.

## 7 Conclusion

Optimizing jet performance proved valuable in improving future BSM search sensitivity. By reconstructing Higgs bosons and charginos more accurately, this paper reveals the importance of determining conditions that best optimize jet performance. Small-R jets were particularly useful at capturing decays whose Higgs bosons were both within the span of large-R jets.

Upon further analysis of small-R jet reconstruction methods, it was found that filtering out events whose jets or Higgs bosons were within the small-R radius significantly improved reconstruction accuracy. In fact, the combined consideration of jet and Higgs boson distances were highly relevant in increasing the accuracy of truth samples across chargino masses: incorrect higher-mass reconstructions disappeared, while the total number of events analyzed did not significantly decrease.

Because chargino mass correlated highly with the accuracy of large-R jet reconstruction, algorithms that further analyze large-R jet reconstruction efficiency would prove to be very useful. Furthermore, combinations of these parameters would be suitable for a neural network to analyze, determining which jet type would best reconstruct a particular decay.

## 8 References

[1] Kane, G (2017). *Modern Elementary Particle Physics (Second Edition)*. Cambridge University Press.

[2] Martin, B and Shaw, G (2001). *Encyclopedia of Physical Science and Technology (Third Edition).* Academic Press Inc.

[3] Salam, G.P., Wang, LT. & Zanderighi, G (2022). The Higgs boson turns ten. *Nature*, 41–47

[4] Schaefer, L (2019). A Search For Wino Pair Production With B−L R-Parity Violating Chargino Decay To A Trilepton Resonance With The ATLAS Experiment.

[5] Resseguie, E (2019). Electroweak Physics at The Large Hadron Collider with the ATLAS Detector: Standard Model Measurement, Supersymmetry Searches, Excesses, and Upgrade Electronics.

[7] Morrison, J (2020*). Modern Physics with Modern Computational Methods (Third Edition)*, Elsevier.

[8] Ryder, L.H (2006). *Encyclopedia of Mathematical Physics*. VDOC.

[9] Dumitru, S., Herwig, C., & Ovrut, B. A (2019). R-parity violating decays of Bino neutralino LSPs at the LHC. *Journal of High Energy Physics*, 1-49

[10] Aad, G., Abbott, B., Abbott, D. C., Abud, A. A., Abeling, K., Abhayasinghe, D. K., ... & Bandieramonte, M (2021). Search for trilepton resonances from chargino and neutralino pair production in s= 13 TeV p p collisions with the ATLAS detector. *Physical Review D*, 112003

[11] Lancaster University Physics Department (2022). Decay Modes and Branching Ratios.

[12] Badea, A., Fawcett, W. J., Huth, J., Khoo, T. J., Poggi, R., & Lee, L. (2022). Solving combinatorial problems at particle colliders using machine learning. Physical Review D, 106(1), 016001.

[13] ATLAS Collaboration. Identification of boosted Higgs bosons decaying into b-quark pairs with the ATLAS detector at 13 TeV. *Eur Phys. J.*

[14] Aad, G., Abbott, B., Abbott, D.C. *et al* (2021). Optimisation of large-radius jet reconstruction for the ATLAS detector in 13 TeV proton–proton collisions. *Eur. Phys. J. C*, 334

[15] ATLAS Collaboration (2022). Searches for exclusive Higgs and $ Z $ boson decays into a vector quarkonium state and a photon using $139 $ fb $^{-1} $ of ATLAS $\sqrt {s}= 13$ TeV proton $-$ proton collision data. *arXiv preprint arXiv:2208.03122*

[16] ATLAS Collaboration., Aad, G., Abbott, B. *et al* (2020). Search for electroweak production of charginos and sleptons decaying into final states with two leptons and

missing transverse momentum in s√=13 TeV *pp* collisions using the ATLAS detector. *Eur. Phys. J. C*, 123

[17] ATLAS Collaboration(2022). Search for resonant $ WZ\rightarrow\ell\nu\ell^{\prime}\ell^{\prime} $ production in proton $-$ proton collisions at $\mathbf {\sqrt {s}= 13} $ TeV with the ATLAS detector. *arXiv preprint arXiv:2207.03925*

[18] Jackson, B (2015). A Search for B-L R-Parity-Violating Scalar Top Decays in √s = 8 TeV pp Collisions With The ATLAS Experiment.

[19] ATLAS Collaboration (2012). Observation of a new particle in the search for the Standard Model Higgs boson with the ATLAS detector at the LHC. *Physics Letters B* **716**, 1-29

[20] CERN (2016). SM Higgs Branching Ratios and Total Decay Widths (update in CERN Report4 2016), *LHCPhysics Web*

[21] The ATLAS Collaboration (2019). Measurement of the photon identification efficiencies with the ATLAS detector using LHC Run 2 data collected in 2015 and 2016. *Eur. Phys. J. C*, 205